\documentclass[%
 reprint,
 amsmath,amssymb,
 aps,
floatfix,
]{revtex4-2}

\usepackage{graphicx}
\usepackage{dcolumn}
\usepackage{bm}
\usepackage{physics}
\usepackage{xcolor}
\usepackage{gensymb}
\usepackage{standalone}
\usepackage[caption=false]{subfig}

\begin{document}

\title{Interfacial Magnon-Mediated Superconductivity in Twisted Bilayer Graphene}

\author{Bj\o rnulf Brekke, Asle Sudb\o\ and Arne Brataas}
\affiliation{Center for Quantum Spintronics, Department of Physics, Norwegian University of Science and Technology, NO-7491 Trondheim, Norway}

\date{\today}

\begin{abstract}
The interfacial coupling between electrons and magnons in adjacent layers can mediate an attractive electron-electron interaction and induce superconductivity. We consider magic-angle twisted bilayer graphene sandwiched between two ferromagnetic insulators to optimize this effect. As a result, magnons induce an interlayer superconducting state characterized by $p$-wave symmetry. We investigate two candidate ferromagnets. The van der Waals ferromagnet CrI$_3$ stands out because it allows compression to tune the superconducting state with an exponential sensitivity. This control adds a new dimension to the tunability of twisted bilayer graphene. Our results open a new path for exploring magnon-induced superconductivity.
\end{abstract}

\maketitle



Heterostructures of ferromagnets (FM) and conductors are currently attracting considerable attention in spintronics. The interfacial coupling between the localized spins and itinerant electrons gives rise to intriguing phenomena such as RKKY interactions, spin-transfer, and spin-pumping \cite{bruno1995theory,Brataas:NatMat2012}.
Furthermore, the coupling between electrons and magnons can mediate an attractive electron-electron interaction \cite{rohling2018superconductivity,hugdal2018,fjaerbu2019superconductivity}. This effect is analogous to the electron-phonon coupling in conventional Bardeen-Cooper-Schrieffer (BCS) superconductivity \cite{bardeen1957theory}.
Superconductivity mediated by magnons has been studied experimentally in different materials \cite{saxena2000superconductivity, aoki2001coexistence, pfleiderer2001coexistence}. Furthermore, superconductivity mediated by (antiferromagnetic) magnons might also appear in certain high-$T_c$ superconductors \cite{si2016high, lee2006doping}.

Superconductivity induced by interfacial coupling to magnons could exist in various material combinations. Examples are normal metals coupled to ferro- and antiferromagnets \cite{rohling2018superconductivity, fjaerbu2019superconductivity}, as well as ferromagnets and antiferromagnets coupled to the surface of topological insulators \cite{kargarian2016amperean, hugdal2018,erlandsen2020magnon,thingstad2021}. The ferromagnetic case has also been experimentally studied \cite{gong2015possible, gong2017time}, showing a superconducting state with a critical temperature $T_c$ significantly higher than the intrinsic superconductivity of two materials. These studies consider either surface effects or monolayer conductors.

Systems designed for interfacial magnon-mediated superconductivity require specific properties. In general, superconducting critical temperatures $T_c$ are exponentially sensitive to interaction strength. In the present case, this is the magnon-mediated electron-electron interaction. Therefore, the electron states should be localized at the interface. Thus, the conducting layer should be as thin as possible yet stable. Furthermore, the electron density of states (DOS) should be large at the Fermi level. In fulfilling these conditions, twisted bilayer graphene (TBG) stands out as an ideal candidate \cite{andrei2020graphene}. 

Twisted bilayer graphene is a two-dimensional material. This renders the electron-magnon-induced effects in TBG more robust than in 3D normal metals, where interactions are constrained to the surface. The relative twist angle of the two graphene layers creates a long-period moiré pattern that, in turn, gives rise to flat electronic bands at certain magic angles \cite{li2010observation, morell2010flat, bistritzer2011moire}. Flat bands at the magic angles greatly enhance the electron DOS. TBG is, therefore, a laboratory for studying the transition from weak- to strong coupling superconductivity by tuning the twist angle. Graphene can couple to conventional ferromagnets \cite{haugen2008spin, wei2016strong, wu2017magnetic}. TBG is also an interesting component of van der Waals heterostructures such as spin valves \cite{geim2013van, cardoso2018van}. Moreover, it is an intrinsic superconductor with a critical temperature of about $T_c=1.7$ K at half filling \cite{cao2018unconventional}. The underlying mechanism is still under debate, and the explanations range from phonon-mediated superconductivity \cite{lian2019twisted, wu2018theory, peltonen2018mean, choi2018strong} to non-BCS type mechanisms \cite{oh2021evidence, yankowitz2019tuning, lu2019superconductors}. 

\begin{figure}[t!]
\includegraphics[width=1\columnwidth]{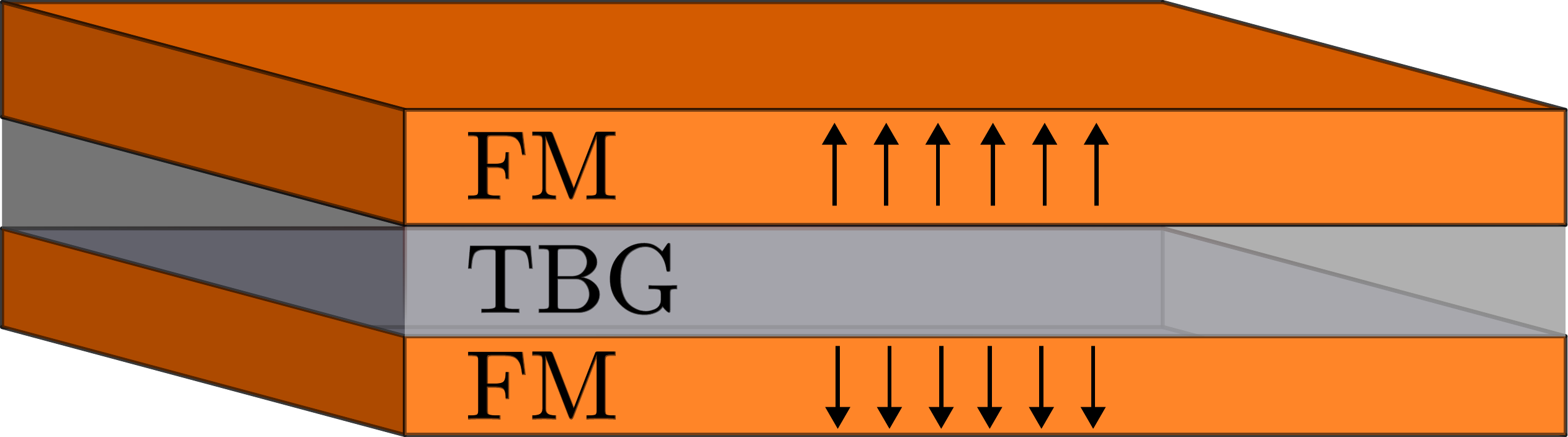}
\caption{\label{fig:heterostructure}We consider a heterostructure consisting of twisted bilayer graphene (TBG) sandwiched between two oppositely aligned ferromagnets (FM).}
\end{figure}

In this Letter, we consider another path to superconductivity in TBG via magnons in adjacent layers. Fig.\ \ref{fig:heterostructure} shows twisted bilayer graphene sandwiched between two identical ferromagnets with oppositely aligned magnetization. The interfacial coupling to magnons gives rise to an effective electron-electron interaction. In combination, TBG´s valley degree of freedom causes a multicomponent superconductor. We find that two of the interlayer coupling channels are suitable for Cooper pair formation. Using a BCS model, we find a $p$-wave superconducting state with a critical temperature of the same order of magnitude as that of the intrinsic mechanism.

To describe the heterostructure, we use a Hamiltonian $H = H_{\mathrm{TBG}} + H_{\mathrm{FM}} + H_{\mathrm{int}}$, where the first term describes the electrons in the twisted bilayer of graphene sheets, and the second term describes the magnons in the top and bottom layer ferromagnets. The last term describes the interfacial coupling between the ferromagnets and the graphene layers. 


We consider a continuum model for the ferromagnets given by
\begin{align}
    H_{\mathrm{FM}} = \int \mathrm{d}^2 \bm{r} \left[  J(\nabla\bm{m})^2 -K_z m_{z}^2 \right)] \, ,
\end{align}
where $\mathbf{m}$ is the magnetization, $J>0$ is the exchange coupling, and the easy-axis anisotropy is parametrized by $K_z>0$. A Holstein-Primakoff transformation to second order in magnon operators and a subsequent Fourier transform yields the magnon Hamiltonian
\begin{align}
\begin{split}
    H_{\mathrm{FM}}=&\sum_{\bm{q}} \omega_{\bm{q}} \left( a^\dagger_{\bm{q}} a_{\bm{q}} + b_{\bm{q}}^\dagger b_{\bm{q}}\right) \, ,
\end{split}
\end{align}
where $a_{\bm{q}}^{(\dagger)}$ and $b_{\bm{q}}^{(\dagger)}$ are the magnon annihilation (creation) operators of momentum $\bm{q}$ in the top and bottom layer, respectively. The magnon dispersion is $\omega_{\bm{q}} = 2JMq^2 + 2MK_z$, where $M$ is the ground state magnetization.

\begin{figure}[ht!]
\includegraphics[width=1\columnwidth]{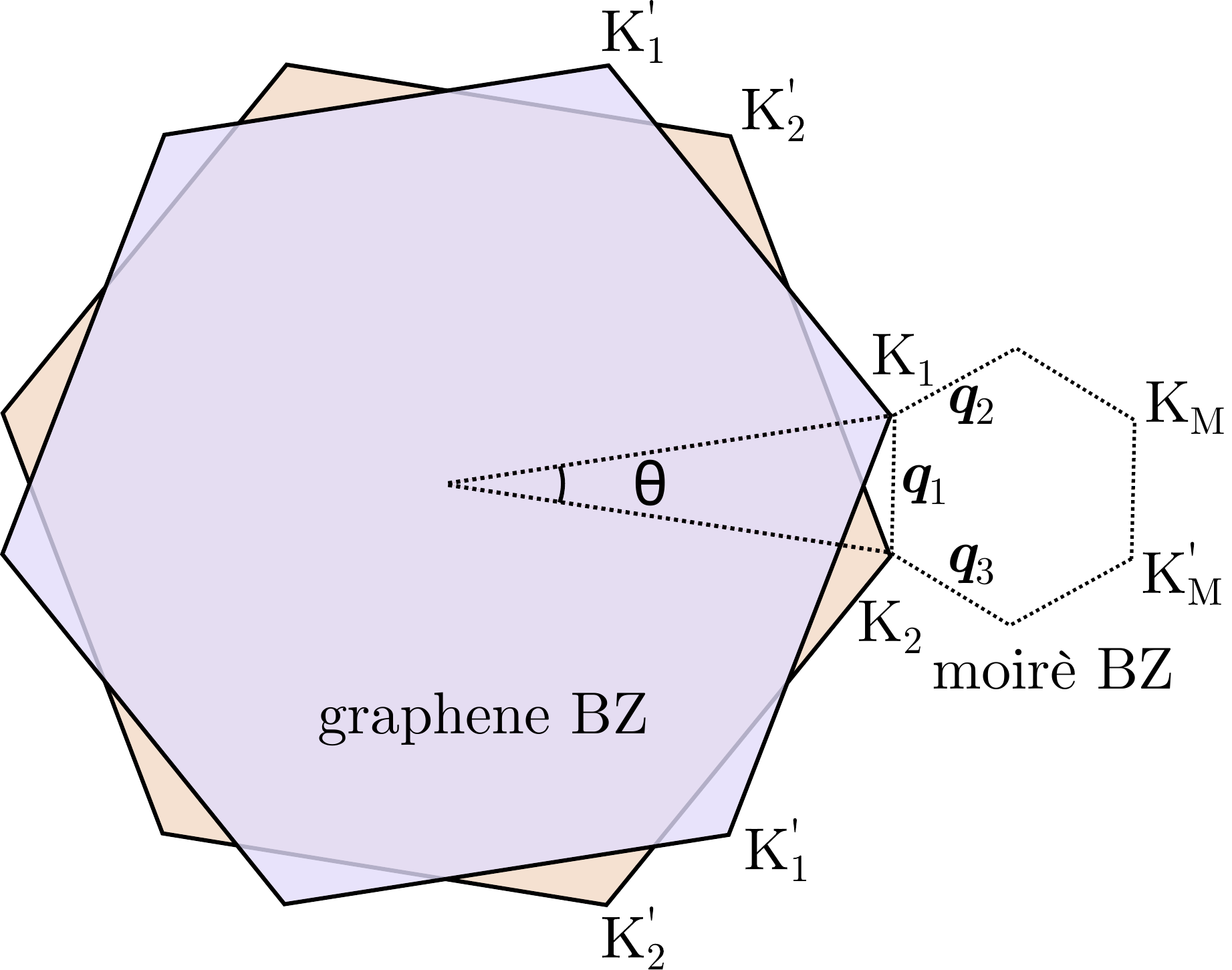}
\caption{\label{fig:BrillouinZone}The Brillouin zones of two graphene layers twisted by an angle $\theta$. These vectors constitute the moirè Brillouin zone with Dirac cones K$_\mathrm{M}$ and K$'_\mathrm{M}$.}
\end{figure}

TBG consists of two graphene monolayers with a relative twist angle $\theta$, as shown in Fig. \ref{fig:BrillouinZone}. In the decoupled limit, the top layer has two Dirac cones $K_1$ and $K'_1$. Similarly, the bottom layer has two Dirac cones $K_2$ and $K'_2$. Because of the relative twist, the Dirac points are related by the three vectors $\bm{q}_1=k_{\theta}(0,-1)^T$, $\bm{q}_2 = k_{\theta}\left(\sqrt{3}/2,1/2\right)^T$ and $\bm{q}_3 = k_{\theta}\left(-\sqrt{3}/2, 1/2\right)^T$. Here, $k_{\theta}=2k_D\sin{\theta/2}$ is the magnitude of $\bm{q}_j$. The coefficient $k_D$ is the magnitude of the Dirac cone momenta in the monolayers. For decoupled layers, the electrons at crystal momentum $\bm{k}$ near $K_1$ or $K_2$ are governed by the single layer Dirac Hamiltonian $h^{K}(\bm{k})=v \bm{\sigma}^*\cdot\bm{k}$, whereas the electrons near the cones $K'_1$ and $K'_2$ are governed by $h^{K'}(\bm{k})=-v \bm{\sigma}\cdot\bm{k}$. Here, $\bm{\sigma}$ is the Pauli matrix vector, and $v$ is the graphene Fermi velocity.

To model the low-energy electron bands of TBG, we use the Bistritzer-MacDonald Hamiltonian \cite{bistritzer2011moire}. In the following, we present the main features required to apply the model for magnon-induced superconductivity. We start with an effective spin-degenerate $4\times 4$ Hamiltonian describing the electrons at cone $K_1$ and interlayer hopping in momentum space to cone $K_2$. It is given by
\begin{align}
    \mathcal{H}_{\bm{k}}^{K_{1} K_2} = \begin{bmatrix}h^{K}(\bm{k}) & T_1 & T_2 & T_3 \\ T_1^\dagger & h^{K}(\bm{k}_1) & 0 & 0 \\ T_2^\dagger & 0 & h^{K}(\bm{k}_2) & 0 \\
    T_3^\dagger & 0 & 0 & h^{K}(\bm{k}_3)\end{bmatrix} \, .
    \label{4x4HamMain}
\end{align}
The three other Hamiltonians describing electrons similarly at the cones $(K_1^{\prime},K_2,K_2^{\prime})$ must also be considered (see Supplemental Material).  
In Eq. \eqref{4x4HamMain}, each element is a $2\times 2$ matrix in sublattice space. The Hamiltonian acts on four two-component spinors $\left(\psi_{K_{1}}(\bm{k}), \psi_{K_2}(\bm{k}_1),  \psi_{K_2}(\bm{k}_2), \psi_{K_2}(\bm{k}_3)\right)^T$. Here, the first spinor component describes electrons near $K_1$, whereas the other three components describe the electrons near $K_2$.
The crystal momentum $\bm{k}_j \equiv \bm{k} - \bm{q}_j$, where $\bm{k}$ is measured relative to the top layer`s Dirac cone $\mathrm{K}_1$.
The Hamiltonian at $K_1$ couples to the Dirac cone $K_2$ through three hopping processes.
The interlayer hopping is captured by the hopping matrices $T_1=w(\sigma_0 + \sigma_x)$, $T_2=w\left(\sigma_0 - 1/2\sigma_x - \sqrt{3}/2\sigma_y\right)$ and $T_3 = w\left(\sigma_0 - 1/2\sigma_x - \sqrt{3}/2\sigma_y\right)$. The hopping strength $w\approx113$ meV \cite{jung2014ab}.

The Hamiltonian in Eq. \eqref{4x4HamMain} exhibits low-energy eigenstates
\begin{align}
    \Psi^{K_1 K_2}=\begin{pmatrix} 1, & -h_1^{-1}T_1^\dagger , & -h_2^{-1}T_2^\dagger, & -h_3^{-1}T_3^\dagger \end{pmatrix}^T,
    \label{state}
\end{align}
where we used the shorthand notation $h_j = h^K(-q_j)$.
Projecting the Hamiltonian on these low-energy states, we get an effective $2\times 2$ sublattice space Hamiltonian on the form
\begin{align}
\begin{split}
    \bra{\Psi^{K_1 K_2}}\mathcal{H}_{\bm{k}}^{K_1 K_2}\ket{\Psi^{K_1 K_2}} = 
     v^{\star}\bm{\sigma}^*\cdot\bm{k}.
    \label{effectiveMain}
\end{split}
\end{align}
This effective Hamiltonian has the form of a single-layer Dirac Hamiltonian with a renormalized velocity
\begin{align}
    \frac{v^{\star}}{v}=\frac{1-3\alpha^2}{1+6\alpha^2} \, ,
\end{align}
where $\alpha=w/ v k_{\theta}$. Note how $\alpha^2=1/3$ yields a vanishing Fermi velocity and flat bands. This value corresponds to the largest "magic angle."

The Hamiltonian in Eq. \eqref{4x4HamMain} models electrons at cone $K_1$ in the top layer and three allowed interlayer hoppings to $K_2$. As noted above, the related Hamiltonians for the $K_1'$ point in the top layer and $K_2$ and $K_2'$ in the bottom layer must also be considered (see Supplemental Material). 
Diagonalizing all four effective $2\times 2$ Dirac Hamiltonians, we find the resulting Hamiltonian
\begin{align}
    \tilde{H}_n^{I, s}(\bm{k}) =  \varepsilon_{n\bm{k}} c^{\dagger}_{\bm{k}nI s}c_{\bm{k}nI s},
\end{align}
with a linear electron dispersion $\varepsilon_{n\bm{k}} = n v^\star \abs{\bm{k}}$.
The $c$-operators are creation- and annihilation operators for the upper $n=1$ and lower  $n=-1$ bands. The eigenstates are superpositions of states at Dirac cones in both layers. Each state is "based" at one of the four cones $I\in \{K_1, K_1', K_2, K'_2$\} with a threefold contribution from the opposing layer. The relative weight of the contributions depends on the twist angle $\theta$ and interlayer hopping strength $w$. We index the eigenstates according to the cone at which it is based. Furthermore, $s$ is the spin index, and the crystal momentum $\bm{k}$ is taken with respect to cone $I$.

\begin{figure}[ht!]
\includegraphics[width=1\columnwidth]{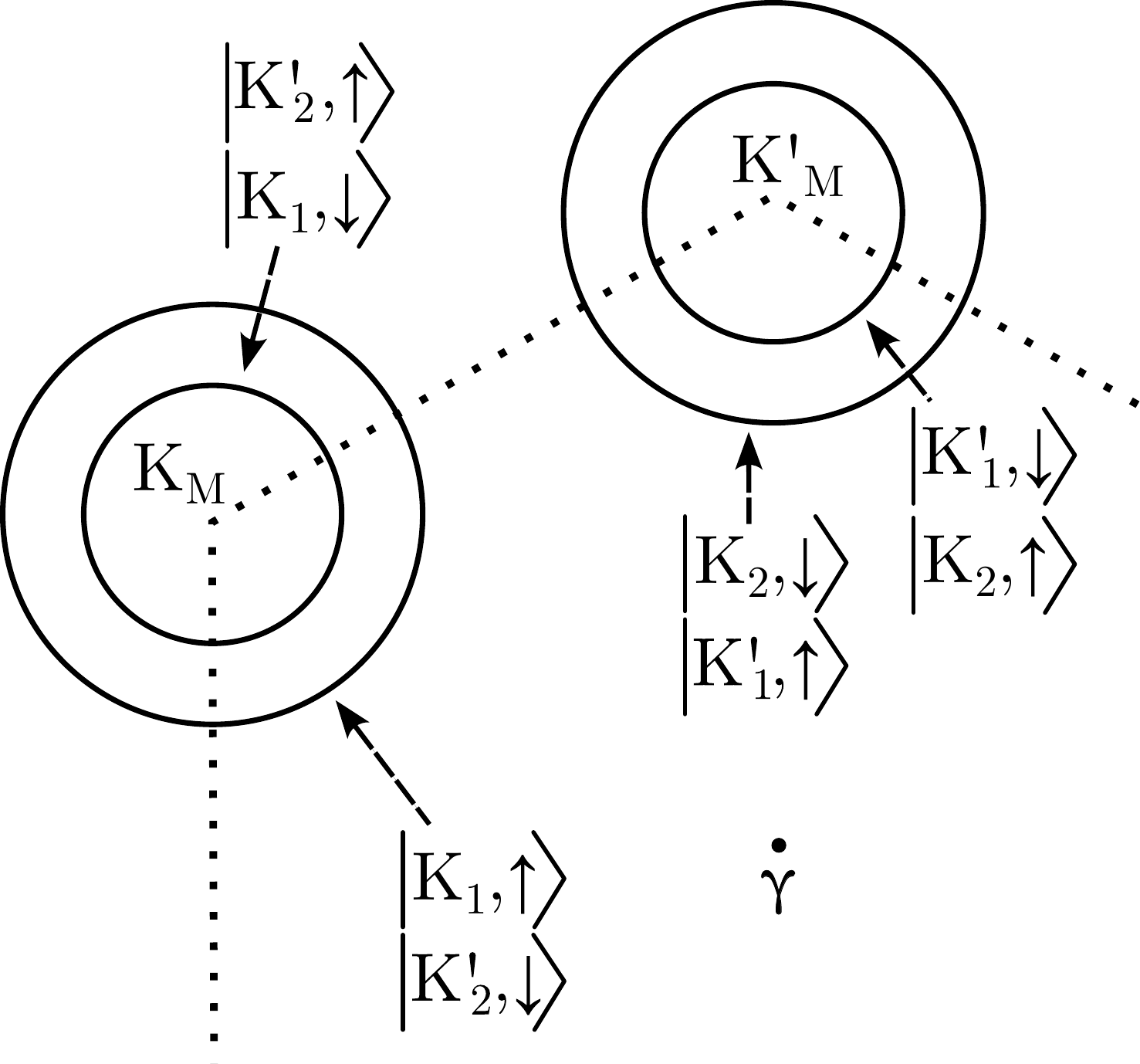}
\caption{\label{fig:moireBZDispersionSplit}An intersection of the electronic bands at constant energy in the moirè Brillouin zone, with moirè cones $K_M$ and $K'_M$ relative to the center $\gamma$. The bands are split due to the interfacial s-d coupling giving rise to Eq. \eqref{SpinSplittingMain}. The bands are doubly degenerate with states labeled according to the cone at which it is based and spin.}
\end{figure}


We model the interfacial coupling with a conventional s-d Hamiltonian
\begin{align}
    H_{\mathrm{int}} = J_{\mathrm{s-d}} \int \mathrm{d}^2\bm{r} \bm{m}(\bm{r})\cdot \bm{s}(\bm{r}),
    \label{startSDMain}
\end{align}
where $\bm{s}$ is the spin operator of the itinerant electrons and $J_{\mathrm{s-d}}$ is the coupling strength. 

As explained above, we take into account spin fluctuations by means of a Holstein-Primakoff transformation. 
We only consider the interfacial coupling between the ferromagnets and their nearest graphene layer. For instance, the first component of the electron state in Eq. \eqref{state} couples to the top ferromagnetic layer, whereas the three next components couple to the bottom layer. The interfacial coupling gives rise to a layer-dependent spin splitting
\begin{align}
    H_{\mathrm{ss}} = \Delta_{ss}\sum_{\bm{k},I,s}ls\left(c_{\bm{k}I s}^{\dagger}c_{\bm{k}I s}\right),
    \label{SpinSplittingMain}
\end{align}
where the spin splitting is given by
\begin{align}
    \Delta_{ss}=\frac{J_{\mathrm{s-d}}M(1-6\alpha^2)}{1+6\alpha^2}.
\end{align}
The layer index $l$ can be extracted directly from $I$. It takes the value $l=1$ for $I\in \{K_1, K'_1\}$ and $l=-1$ for $I \in \{K_2, K'_2\}$. The spin index $s=1$ for spin up and $s=-1$ for spin down. 
The spin-split electron dispersion in the moirè Brillouin zone is shown in Fig. \ref{fig:moireBZDispersionSplit}. In the case of a monolayer conductor sandwiched between two oppositely aligned ferromagnets, the spin splitting vanishes exactly. This is not necessarily the case for a bilayer, as seen from Eq. \eqref{SpinSplittingMain}. Each of the electron eigenstates is asymmetrically delocalized in the two layers. Hence, the net magnetic field shifts the energy of the electron states. The sign of the shift depends on spin and the layer in which the state is "based." At the special twist angle $\alpha^2 = 1/6$, the states are symmetrically distributed in the two layers such that the spin shift cancels. 

We next consider the electron-magnon coupling
\begin{align}
\begin{split}
    H_{\mathrm{e-m}} =& \sum_{\bm{q},\bm{k},I} \left[ g_{I,\bm{k},\bm{q}}^{a} \left(c_{\bm{k}+\bm{q}I\uparrow}^\dagger c_{\bm{k}I\downarrow} \right)a_{\bm{q}} + h.c. \right. \\ &+ \left. g_{I,\bm{k},-\bm{q}}^{b} \left(c_{\bm{k}-\bm{q}I\uparrow}^\dagger c_{\bm{k}I\downarrow} \right)b^{\dagger}_{\bm{q}} + h.c. \right] \, . 
\end{split}
\end{align}
The coupling parameters $g_{I,\bm{k},\bm{q}}^{a}$ and $g_{I,\bm{k},-\bm{q}}^{b}$ depend on the index $I$, in addition to the ferromagnetic layers $a$ and $b$ to which it couples. Their full form is given in the Supplemental material.


We derive an effective electron-electron interaction between electrons in state $I$ and $I'$ via a Schrieffer-Wolff transformation \cite{schrieffer1966relation}, obtaining 
\begin{align}
    H_{\mathrm{eff}}=\sum_{\bm{k},\bm{k}',I,I'} V_{\bm{k}\bm{k}' II'}c_{\bm{k}'I\uparrow}^\dagger c_{-\bm{k}'I'\downarrow}^\dagger c_{-\bm{k}I\downarrow}  c_{\bm{k}I'\uparrow},
    \label{InteractionTerm}
\end{align}
where the interaction strength $V_{\bm{k}\bm{k}' II'}$ is given by
\begin{align}
\begin{split}
    V_{\bm{k}\bm{k}' II'} =  \frac{2\omega_{\bm{k'+k}}\left(\sum_{j\in \{a,b\}}g_{I,-\bm{k},\bm{k}'+\bm{k},}^{j} g_{I',-\bm{k},\bm{k}'+\bm{k}}^{j}\right)}{\omega^2_{\bm{k}'+\bm{k}}-(\varepsilon_{\bm{k}'}-\varepsilon_{\bm{k}})^2}.
    \label{effectiveInteractionMain}
\end{split}
\end{align}
The rich $I, I'$ structure of Eq. \eqref{effectiveInteractionMain} yields, in principle, a large number of possible pairing channels. However, not all of them are suitable for the formation of Cooper pairs. For instance, states based in the same layer with opposite spins are spin split and thus not suitable for spin-unpolarized Cooper-pairing either in the spin-singlet or spin-triplet channels. Spin-unpolarized Cooper-pairing is the only possibility when the electron-magnon coupling originates from collinear spin ground states in the magnetic insulator \cite{Maeland_arxiv_2022}. Hence, we focus on the coupling between electrons in different layers. This excludes half of the pairing channel candidates. Furthermore, requiring the Cooper pairs to have a zero net momentum with respect to the moiré Brillouin zone leaves two possible pairing channels. These are inter-layer intra-valley pairs denoted by
\begin{align}
    \left\{I = K_1,\bar{I} = K_2\right\}, &&
    \left\{I = K'_1, \bar{I} = K'_2\right\}.
    \label{PairingChannelMain}
\end{align}
Within these pairing channels, the interaction can be further decomposed in terms of pairing symmetry. To that end, we approximate the magnon frequency to be constant $\omega_{\bm{k}'+\bm{k}}=\omega_M$. In this approximation, the full angular dependence originates from the coupling constants $g$ in Eq. \eqref{effectiveInteractionMain}. The effective interaction decomposes into an $s$-, a $p$-, and a $d$-wave component. In the low-frequency limit $(\varepsilon_{\bm{k}'}-\varepsilon_{\bm{k}})^2 \rightarrow 0$,  the $s$- and $d$-wave components are repulsive. In contrast, the $p$-wave symmetry component is attractive and enables Cooper pair formation. Hence, we expect TBG to exhibit an interlayer magnon-mediated superconducting state with $p$-wave symmetry.

We now give an estimate of the critical temperature from the conventional BCS expression
\begin{align}
    k_B T_c  \approx 1.13\omega_M e^{-\frac{1}{\lambda}},
\label{BCS-Tc}
\end{align}
where $\omega_M$ is the characteristic magnon frequency. The coupling $\lambda = V N'_D$ depends on the effective interaction $V$ and the DOS $N'_D$ per valley per layer per spin. The electron DOS is enhanced near the magic angle due to the flat energy bands. Ref. \cite{carr2019exact} reports a total DOS $N_D \gtrapprox 100 \; \mathrm{eV}^{-1} \mathrm{nm}^{-2}$ close to the magic angle. This suggests a DOS $N_D' = 12.5\; \mathrm{eV}^{-1} \mathrm{nm}^{-2}$. Although the BCS theory does not predict the critical temperature, it may be used to obtain estimates of $T_c$. An important feature of Eq. \eqref{BCS-Tc} is the non-perturbative renormalization of the magnon energy scale, in that $T_c$ depends exponentially on the inverse of the DOS and the interfacial s-d coupling.


The repulsive Coulomb interaction can be detrimental to the superconducting state. TBG´s Coulomb interaction is largely screened at the magic angle for long wavelengths due to a large twist-angle dependent dielectric constant $\epsilon > 250$ \cite{Goodwin2019PRB}. The Coulomb coupling strength is $\mu = V_C(k_{\theta})N_D \approx 1$. It slightly exceeds the attractive magnon-mediated interaction. However, for two reasons, the superconducting state is robust in the presence of this Coulomb repulsion. First, the superconducting gap function is of interlayer $p$-wave symmetry. Hence, the Cooper pairs circumvent the significant on-site $s$-wave contribution of the Coulomb interaction. However, we will not consider the decomposition of the Coulomb interaction as the approximation is already at a crude level. Second, the Coulomb interaction is frequency independent at the scale of the magnon cut-off frequencies. To account for this, we adopt the Morel-Anderson model \cite{Morel-Anderson} to find an effective coupling strength
\begin{align}
    \mu^* = \frac{\mu}{1+\mu\ln{\frac{\omega_p}{\omega_M}}} \approx 0.13.
\end{align}
Here, we used the observed interband plasmon frequency $\omega_p \approx 200 \; \mathrm{meV}$ as the Coulomb interaction cut off \cite{hesp2021observation}. The effective interaction strength takes the value $\lambda^* = \lambda - \mu^*$.

Interfacial coupling between graphene and ferromagnets has been studied theoretically and experimentally for numerous materials \cite{swartz2012integration,averyanov2018high,wei2016strong,wang2015proximity,farooq2019switchable,zhang2018strong,holmes2020exchange}. Hence, there are several candidates. Here, we consider two specific ferromagnets.

EuO is a ferromagnetic semiconductor with Curie temperature $T_c = 69$ K. It has an fcc unit cell with lattice constant $5.1$ Å. Hence, two magnetic Eu$^{2+}$ ions per unit cell, each with spin $S=7/2$, are located at the interface and thus accessible for interfacial s-d coupling \cite{mauger1986magnetic}. EuO thin films can be deposited on graphene epitaxially \cite{swartz2012integration, averyanov2018high}. The induced exchange splitting is found to be $\Delta=36$ meV \cite{yang2013proximity}. At the wave number $k_{\theta}$, $\omega_M = 0.15$ meV is an appropriate frequency cut-off. These parameters suggest an effective coupling strength $\lambda^* \approx 0.8$ and a critical temperature $T_C \approx 0.5$ K.

CrI$_3$ is a van der Waals ferromagnet down to the monolayer limit \cite{huang2017layer}. The crystal has two magnetic ions per unit cell. Each magnetic ion carries a magnetic moment $S=3\mu_B$ \cite{mcguire2015coupling}. CrI$_3$ hosts two magnonic modes accessible for electron-magnon coupling. Their respective energies at momentum $\bm{q}=0$ are $0.3$ meV and $17$ meV \cite{cenker2021direct}. In a graphene-CrI$_3$ heterostructure, CrI$_3$ is theoretically found to induce an exchange splitting of $20$ meV \cite{holmes2020exchange}.
Considering CrI$_3$ as the ferromagnet, we find a coupling constant $\lambda^* \approx 0.4$ and critical temperature $T_c \approx 0.3$ K.

Van der Waals ferromagnets, such as CrI$_3$, are particularly interesting candidates.
This is because the interfacial exchange splitting increases significantly under compression.
For CrI$_3$, a slight decrease in the interlayer gap can enhance the exchange splitting. A moderate reduction of the interlayer distance $\Delta d = -0.5$ Å leads to an exchange splitting of 80 meV. The splitting can reach values up to 150 meV \cite{zhang2018strong}. The enhanced interfacial interaction renders the higher energy magnon branch of CrI$_3$ accessible for electron-magnon coupling. We expect this to increase the critical temperature significantly. In this way, the van der Waals spin valve exhibits a tunable compression-controlled superconducting state that connects the weak- and strong-coupling regimes. Due to the limited validity of the weak-coupling BCS model, we do not estimate the critical temperature for the compressed heterostructure.



In conclusion, we have demonstrated that interfacial magnons can induce superconductivity in twisted bilayer graphene. The magnons yield a multicomponent superconducting state due to the valley structure of TBG. We find an attractive interlayer channel suitable for Cooper pairs with $p$-wave symmetry. Moreover, we have considered two promising candidate ferromagnets, EuO and CrI$_3$. Both exhibit a critical temperature of the same order of magnitude as the intrinsic superconducting mechanism. Using van der Waals magnets is particularly interesting because the interlayer interaction strength is tunable through compression. For this reason, the superconducting state is tunable both via the twist angle and external compression. Twisted bilayer graphene sandwiched between ferromagnets is, therefore, a promising platform in which to explore magnon-mediated superconductivity.

The Research Council of Norway (RCN) supported this work through its Centres of Excellence funding scheme, project number 262633, "QuSpin", as well as RCN project number 323766.




\bibliography{apssamp}
\pagebreak
\widetext
\begin{center}
\textbf{\large Supplemental material to "Interfacial Magnon-Mediated Superconductivity in Twisted Bilayer Graphene"}
\end{center}
\setcounter{equation}{0}
\setcounter{figure}{0}
\setcounter{table}{0}
\setcounter{page}{1}
\makeatletter
\renewcommand{\theequation}{S\arabic{equation}}
\renewcommand{\thefigure}{S\arabic{figure}}

\onecolumngrid
\section{Ferromagnet\label{Ferromagnet}}
We consider a continuum model of magnetization dynamics in a ferromagnetic layer. The Hamiltonian is 
\begin{align}
    H_m = \int \mathrm{d}^2\bm{r} \left[ J(\nabla\bm{m})^2 -K_z m_z^2 \right] \, ,
\end{align}
where the exchange constant $J>0$, the easy-axis anisotropy $K_z>0$ and $\bm{m}$ is the magnetization. We use this model for both the top and bottom ferromagnetic layers introduced in the main text.
We employ the Holstein-Primakoff transformation for the variables
\begin{align}
    m_x = \frac{1}{2}(m^+ + m^-) \, , && 
    m_y = \frac{1}{2i}(m^+ - m^-) \, , && m_z = m_z \, . 
\end{align}
We quantize the magnetization to quadratic order in the bosonic operators for the top layer using
\begin{align}
    m^{+}=\sqrt{2M}a_{\bm{r}} \, , && m^{-}=\sqrt{2M}a_{\bm{r}}^\dagger \, , &&
    m^{z} = (M -a_{\bm{r}}^\dagger a_{\bm{r}}) \, .
    \label{HPtop}
\end{align}
The ground state magnetization $M$ of the bottom layer is oppositely aligned. Hence, we quantize the operators as
\begin{align}
    m^{+}=\sqrt{2M}b^{\dagger}_{\bm{r}} \, , && m^{-}=\sqrt{2M}b_{\bm{r}} \, , &&
    m^{z} = -(M - b_{\bm{r}}^\dagger b_{\bm{r}}) \, .
    \label{HPbottom}
\end{align}
We consider the Hamiltonian in momentum space by using the Fourier transforms
\begin{align}
    a_{\bm{r}} = \frac{1}{\sqrt{\mathcal{V}}}\sum_{\bm{k}}e^{i\bm{k}\cdot\bm{r}}a_{\bm{k}}, && b_{\bm{r}} = \frac{1}{\sqrt{\mathcal{V}}}\sum_{\bm{k}}e^{i\bm{k}\cdot\bm{r}}b_{\bm{k}},
\end{align}
where $\mathcal{V}$ is the area of the ferromagnet. The transformations yield the magnon Hamiltonian
\begin{align}
\begin{split}
    H_m=&\sum_{\bm{q}}  \omega_{\bm{q}} \left( a^\dagger_{\bm{q}} a_{\bm{q}} + b_{\bm{q}}^\dagger b_{\bm{q}}\right) \, ,
\end{split}
\end{align}
with $\omega_{\bm{q}} = 2JMq^2 + 2MK_z$ as stated in the main text. Here, and throughout the text, we set $\hbar=1$.

\section{Bistritzer-MacDonald model for TBG\label{BM}}
A single sheet of graphene has two Dirac cones in the first Brillouin zone located at the high-symmetry points $K$ and $K'$. These cones are related by inversion symmetry and are described by Dirac Hamiltonians of opposite chirality
\begin{align}
    h^K(\bm{k}) =  v \bm{\sigma}^*\cdot\bm{k} \, , && h^{K'}(\bm{k}) = - v \bm{\sigma}\cdot\bm{k} \, .
\end{align}
Here, $\bm{\sigma}$ is the vector of the Pauli matrices acting on the sublattice space. In the bilayer case, we use $K_1$ and $K_1'$ for the top layer and $K_2$ and $K_2'$ for the bottom layer.
The Bistritzer-MacDonald (BM) model describes interlayer hopping between these Dirac cones in twisted bilayer graphene \cite{bistritzer2011moire}. The hopping from momentum $\bm{k}$ relative to the cone $K_1$ in the top layer to the bottom layer cone $K_2$ is described by the Hamiltonian
\begin{align}
    \mathcal{H}_{\bm{k}}^{K_1 K_2} = \begin{bmatrix}h^{K}(\bm{k}) & T_1 & T_2 & T_3 \\ T_1^\dagger & h^{K}(\bm{k}-\bm{q}_1) & 0 & 0 \\ T_2^\dagger & 0 & h^{K}(\bm{k}-\bm{q}_2) & 0 \\
    T_3^\dagger & 0 & 0 & h^{K}(\bm{k}-\bm{q}_3)\end{bmatrix}.
    \label{4x4Ham}
\end{align}
Here, each element is a $2\times 2$ matrix with respect to sublattice space. The $\bm{q}$ vectors are defined as
\begin{align}
    \bm{q}_1 = k_{\theta}(0,-1)^T, && \bm{q}_2 = k_{\theta}\left(\frac{\sqrt{3}}{2},\frac{1}{2}\right)^T && \bm{q}_3 = k_{\theta}\left(-\frac{\sqrt{3}}{2},\frac{1}{2}\right)^T,
    \label{qVectors}
\end{align}
where the $\bm{q}$-vector magnitude is $k_{\theta}=2k_D\sin{\theta/2}$. The factor $k_D$ is the magnitude of the Dirac cone momenta in the single layers. The interlayer hopping matrices are $T_1=w(\sigma_0 + \sigma_x)$, $T_2=w\left(\sigma_0 - 1/2\sigma_x - \sqrt{3}/2\sigma_y\right)$ and $T_3 = w\left(\sigma_0 - 1/2\sigma_x - \sqrt{3}/2\sigma_y\right)$, where $w$ is the interlayer hopping strength.

The Hamiltonian acts on the four two-component spinors $\left(\psi_{K_{1}}(\bm{k}), \psi_{K_2}(\bm{k}_1),  \psi_{K_2}(\bm{k}_2), \psi_{K_2}(\bm{k}_3)\right)^T$. Each component is a spinor in the sublattice space. The first component $\psi_{K_{1}}(\bm{k})$ is located near $K_1$ whereas the three components $\psi_{K_2}(\bm{k}_1)$,  $\psi_{K_2}(\bm{k}_2)$ and  $\psi_{K_2}(\bm{k}_3)$ are located in the bottom layer near $K_2$ at the three distinct momenta $\bm{k}_j = \bm{k} - \bm{q}_j$. The crystal momentum $\bm{k}$ is relative to the valley $K_1$. In this basis, we project the Hamiltonian on the two-fold low-energy eigenstate
\begin{align}
    \bm{\Psi}^{K_1 K_2}=\begin{pmatrix} 1, & -h_1^{-1}T_1^\dagger , & -h_2^{-1}T_2^\dagger, & -h_3^{-1}T_3^\dagger \end{pmatrix}^T,
    \label{eigenstate}
\end{align}
where $h_j = h^K(-\bm{q}_j)$.
The state is normalized to
\begin{align}
    \abs{\bm{\Psi}^{K_1 K_2}}^2 = 1 + 6\alpha^2.
\end{align}
This is evident from the identity
\begin{align}
    h_1^{-1}T_1^\dagger T_1 h_1^{-1} + h_2^{-1}T_2^\dagger T_2 h_2^{-1} + h_3^{-1}T_3^\dagger T_3 h_3^{-1} = 6\alpha^2\sigma_0.
\end{align}
Here, $\sigma_0$ denotes the $2\times 2$ identity matrix.
The parameter $\alpha=w/ v k_{\theta}$, where $v$ is the single layer graphene Fermi velocity. 
Direct application of the Hamiltonian on the state $\Psi^{K_1 K_2}$ in Eq. \eqref{4x4Ham} shows that the state is a zero eigenstate at $\bm{k}=0$ and has a linear dispersion with respect to $\bm{k}$. More precisely, the projected low-energy effective Hamiltonian is
\begin{align}
    \bra{\bm{\Psi}^{K_1 K_2}}\mathcal{H}_{\bm{k}}^{K_1 K_2}\ket{\bm{\Psi}^{K_1 K_2}} = \frac{ v}{1+6\alpha^2}\left[\bm{\sigma}^*\cdot\bm{k} + \sum_{j=1}^3 T_j h_j^{-1\dagger}\bm{\sigma}^*\cdot\bm{k}h_j^{-1}T_j^\dagger\right]  =
     v^{\star}\bm{\sigma}^*\cdot\bm{k},
    \label{effective}
\end{align}
where
\begin{align}
    \frac{v^{\star}}{v}=\frac{1-3\alpha^2}{1+6\alpha^2}
\end{align}
shows the twist-angle dependence of the renormalized Fermi velocity $v^{\star}$.

\subsection{The full Hamiltonian\label{FullBM}}
So far, we have modeled the interlayer hopping from cone $K_1$ to cone $K_2$. In total, there are four inequivalent Dirac cones, two in each layer. In this section, we argue by symmetry to find the Hamiltonians for the three other interlayer hopping processes. We omit the spin quantum number in this section because the interlayer hopping and the Dirac Hamiltonian are spin independent.

In general, the total Hamiltonian
\begin{align}
    H_{\bm{k}} = \mathcal{H}_{\bm{k}}^{K_1 K_2} \oplus \mathcal{H}_{\bm{k}}^{K'_1 K'_2} \oplus \mathcal{H}_{\bm{k}}^{K_2 K_1} \oplus \mathcal{H}_{\bm{k}}^{K'_2 K'_1},
\end{align}
must be invariant under the symmetries of the twisted bilayer. We relate the terms of the Hamiltonian by considering symmetry transformations that relate the cones. Ref. \cite{lian2019twisted} outlines a similar procedure.

\subsection{Time-reversal symmetry}

We start by considering time-reversal symmetry $\mathcal{T}$. It relates primed and unprimed Dirac cones, such that $K\leftrightarrow K'$. Furthermore, it acts as complex conjugation on the sublattice Pauli matrices.
The hopping matrices transform in the following way,
\begin{align}
    T_2^{(\dagger)} \leftrightarrow T_3^{(\dagger)},
\end{align}
whereas $T_1^{(\dagger)}$ is left invariant. The crystal momentum $\bm{k}\rightarrow -\bm{k}$.
In total, we have that
\begin{align}
    \mathcal{H}_{\bm{k}}^{K'_1 K'_2} = \begin{bmatrix}h^{K'}(\bm{k}) & T_1 & T_3 & T_2 \\ T_1^\dagger & h^{K'}(\bm{k}+\bm{q}_1) & 0 & 0 \\ T_3^\dagger & 0 & h^{K'}(\bm{k}+\bm{q}_2) & 0 \\
    T_2^\dagger & 0 & 0 & h^{K'}(\bm{k}+\bm{q}_3)\end{bmatrix}.
    \label{4x4HamTime}
\end{align}

\subsection{$C_{2x}$ rotation symmetry}
We now use the twofold $C_{2x}$ symmetry to find $H^{K_2 K_1}$ and $H^{K_2' K'_1}$. This symmetry exchanges the sublattices and flips the sign of the $ k_y$ component. Hence,
\begin{align}
    \left(\sigma_x, \sigma_y\right) \rightarrow \left(\sigma_x, -\sigma_y\right), && \left(k_x, k_y\right) \rightarrow \left(k_x, -k_y\right). 
\end{align}
Additionally, $C_{2x}$ exchanges the top and bottom layers. In total,
\begin{align}
    \mathcal{H}_{\bm{k}}^{K_2 K_1} = \begin{bmatrix}h^K(\bm{k}) & T_1 & T_3 & T_2 \\ T_1^\dagger & h^K(\bm{k}+\bm{q}_1) & 0 & 0 \\ T_3^\dagger & 0 & h^K(\bm{k}+\bm{q}_3) & 0 \\
    T_2^\dagger & 0 & 0 & h^K(\bm{k}+\bm{q}_2)\end{bmatrix}.
    \label{4x4HamC2}
\end{align}

\subsection{$C_{2x}\mathcal{T}$ symmetry}
The combination of the symmetries yields
\begin{align}
    \mathcal{H}_{\bm{k}}^{K_2' K'_1} = \begin{bmatrix}h^{K'}(\bm{k}) & T_1 & T_2 & T_3 \\ T_1^\dagger & h^{K'}(\bm{k}-\bm{q}_1) & 0 & 0 \\ T_2^\dagger & 0 & h^{K'}(\bm{k}-\bm{q}_3) & 0 \\
    T_3^\dagger & 0 & 0 & h^{K'}(\bm{k}-\bm{q}_2)\end{bmatrix}.
    \label{4x4HamC2Time}
\end{align}

\subsection{\label{Summary}Summary of full Hamiltonian}
To summarize this section, we list the four distinct Hamiltonians
\begin{subequations}
\begin{align}
    \mathcal{H}_{\bm{k}}^{K_1 K_2} = \begin{bmatrix}h^K(\bm{k}) & T_1 & T_2 & T_3 \\ T_1^\dagger & h^K(\bm{k}-\bm{q}_1) & 0 & 0 \\ T_2^\dagger & 0 & h^K(\bm{k}-\bm{q}_2) & 0 \\
    T_3^\dagger & 0 & 0 & h^K(\bm{k}-\bm{q}_3)\end{bmatrix},
    \label{4x4Ham1}
\end{align}

\begin{align}
    \mathcal{H}_{\bm{k}}^{K_1' K_2'} = \begin{bmatrix}h^{K'}(\bm{k}) & T_1 & T_3 & T_2 \\ T_1^\dagger & h^{K'}(\bm{k}+\bm{q}_1) & 0 & 0 \\ T_3^\dagger & 0 & h^{K'}(\bm{k}+\bm{q}_2) & 0 \\
    T_2^\dagger & 0 & 0 & h^{K'}(\bm{k}+\bm{q}_3)\end{bmatrix},
    \label{4x4Ham2}
\end{align}

\begin{align}
    \mathcal{H}_{\bm{k}}^{K_2 K_1} = \begin{bmatrix}h^K(\bm{k}) & T_1 & T_3 & T_2 \\ T_1^\dagger & h^K(\bm{k}+\bm{q}_1) & 0 & 0 \\ T_3^\dagger & 0 & h^K(\bm{k}+\bm{q}_3) & 0 \\
    T_2^\dagger & 0 & 0 & h^K(\bm{k}+\bm{q}_2)\end{bmatrix},
    \label{4x4Ham3}
\end{align}

\begin{align}
    \mathcal{H}_{\bm{k}}^{K_2' K_1'} = \begin{bmatrix}h^{K'}(\bm{k}) & T_1 & T_2 & T_3 \\ T_1^\dagger & h^{K'}(\bm{k}-\bm{q}_1) & 0 & 0 \\ T_2^\dagger & 0 & h^{K'}(\bm{k}-\bm{q}_3) & 0 \\
    T_3^\dagger & 0 & 0 & h^{K'}(\bm{k}-\bm{q}_2)\end{bmatrix}.
    \label{4x4Ham4}
\end{align}
\label{4x4HamSummary}
\end{subequations}

We can now find the low-energy eigenstates that diagonalize each of the Hamiltonians. For a compact notation, we use
\begin{align}
    \pm h_{j} = h^K(\mp{q}_j), && \pm h'_{j} = h^{K'}(\mp{q}_j).
\end{align}
Inspired by Eq. \eqref{eigenstate}, we read off the normalized eigenstates directly as
\begin{subequations}
\begin{align}
    \bm{\Psi}^{K_1 K_2} &= \frac{1}{\sqrt{1+6\alpha^2}}\left(1, -h_1^{-1}T_1^\dagger,  -h_2^{-1}T_2^\dagger, -h_3^{-1}T_3^\dagger\right)^T, \\
    \bm{\Psi}^{K_1' K_2'} &= \frac{1}{\sqrt{1+6\alpha^2}}\left(1, {h'_1}^{-1}T_1^\dagger,  {h'_2}^{-1}T_3^\dagger, {h'_3}^{-1}T_2^\dagger\right)^T, \\
    \bm{\Psi}^{K_2 K_1} &= \frac{1}{\sqrt{1+6\alpha^2}}\left(1, h_1^{-1}T_1^\dagger,  h_3^{-1}T_3^\dagger, h_2^{-1}T_2^\dagger\right)^T, \\
    \bm{\Psi}^{K_2' K'_1} &= \frac{1}{\sqrt{1+6\alpha^2}}\left(1, -{h'_1}^{-1}T_1^\dagger,  -{h'_3}^{-1}T_2^\dagger, -{h'_2}^{-1}T_3^\dagger\right)^T.
\end{align}
\label{eigenstatesSummary}
\end{subequations}
Each eigenstate is taken with respect to a specific basis, i.e., the basis of the Hamiltonian, to which the eigenstate belong. For completeness, we present the bases as
\begin{subequations}
\begin{align}
    \mathcal{B}^{K_1 K_2} &= \left(\psi_{K_{1}}(\bm{k}), \psi_{K_2}(\bm{k}-\bm{q}_1),  \psi_{K_2}(\bm{k}-\bm{q}_2), \psi_{K_2}(\bm{k}-\bm{q}_3)\right)^T, \\
    \mathcal{B}^{K_1' K_2'} &= \left(\psi_{K'_1}(\bm{k}), \psi_{K'_2}(\bm{k}+\bm{q}_1),  \psi_{K'_2}(\bm{k}+\bm{q}_2), \psi_{K'_2}(\bm{k}+\bm{q}_3)\right)^T, \\
    \mathcal{B}^{K_2 K_1} &= \left(\psi_{K_2}(\bm{k}), \psi_{K_1}(\bm{k}+\bm{q}_1),  \psi_{K_1}(\bm{k}+\bm{q}_3), \psi_{K_1}(\bm{k}+\bm{q}_2)\right)^T, \\
    \mathcal{B}^{K_2' K'_1} &= \left(\psi_{K'_2}(\bm{k}), \psi_{{K'_1}}(\bm{k}-\bm{q}_1),  \psi_{K'_1}(\bm{k}-\bm{q}_3), \psi_{K'_1}(\bm{k}-\bm{q}_2)\right)^T.
\end{align}
\label{bases}
\end{subequations}

\subsection{The effective $2\times2$ electron Hamiltonian\label{2x2Ham}}
In this section, we consider the effective Hamiltonians in Eq. \eqref{4x4HamSummary} with respect to their respective low-energy eigenstates given in Eq. \eqref{eigenstatesSummary}. As seen from Eq. \eqref{effective}, this yields an effective $2 \times 2$ model. To simplify notation, we write only the first of the two superscripts used in Eqs. \eqref{4x4HamSummary} and \eqref{eigenstatesSummary}. Nevertheless, we emphasize that the eigenstates are still superpositions of electron states at cones in distinct layers.
The set of effective Hamiltonians is given by
\begin{align}
    \bra{\bm{\Psi}^{K^{\eta}_l}} \mathcal{H}^{K^{\eta}_l }_{\bm{k}} \ket{\bm{\Psi}^{K^{\eta}_l}} = v^{\star} (\eta \sigma_x k_x - \sigma_y k_y).
\end{align}
Here, we introduced $\eta$ to account for the chirality of the Dirac cones. That is, $\eta = 1$ and $\eta = -1$ corresponds to the Dirac cones $K$ and $K'$, respectively. The subscript $l$ denotes the layer of the cone. The eigenvalues of the effective Hamiltonians are 
\begin{align}
    \varepsilon_{\pm} = \pm  v^\star \abs{\bm{k}},
\end{align}
and the corresponding eigenvectors are
\begin{align}
    \phi_{\bm{k}-}^\eta = \frac{1}{\sqrt{2}}(-\eta e^{\eta\varphi_{\bm{k}}}, 1)^T, && \phi_{\bm{k}+}^\eta = \frac{1}{\sqrt{2}}(\eta e^{\eta \varphi_{\bm{k}}}, 1)^T.
    \label{phiFactors}
\end{align}
Here, $\varphi_{\bm{k}}$ is the angle between $\bm{k}$ and the $x$-axis.
The eigenvectors span a unitary transformation $U_{\bm{k}\eta}$ such that
\begin{align}
     v^\star U_{\bm{k}\eta}^\dagger (\eta \sigma_x k_x -\sigma_y k_y) U_{\bm{k}\eta} = \begin{pmatrix}\varepsilon_{-} & 0 \\ 0 & \varepsilon_{+}\end{pmatrix},
\end{align}
where
\begin{align}
    U_{\bm{k}\eta} = \frac{1}{\sqrt{2}}\begin{pmatrix}-\eta e^{i\eta\varphi_{\bm{k}}} && \eta e^{i\eta \varphi_{\bm{k}}} \\ 1 && 1 \end{pmatrix} && U^\dagger_{\bm{k}\eta} = \frac{1}{\sqrt{2}}\begin{pmatrix}-\eta e^{-i\eta\varphi_{\bm{k}}} && 1 \\ \eta e^{-i\eta \varphi_{\bm{k}}} && 1 \end{pmatrix}
\end{align}
Accordingly, the eigenstates in Eq. \eqref{eigenstatesSummary} are related to band electron operators by the transformations
\begin{subequations}
\begin{align}
    c_{\bm{k}I} = U^\dagger_{\bm{k}\eta} \bm{\Psi}^{I} && c^\dagger_{\bm{k}I} = \bm{\Psi}^{I\dagger}U_{\bm{k}\eta} \\
    \bm{\Psi}^{I} = U_{\bm{k}\eta} c_{\bm{k}I}  &&   \bm{\Psi}^{I\dagger}=  c^\dagger_{\bm{k}I}U^\dagger_{\bm{k}\eta}
\end{align}
\label{UTransform}
\end{subequations}
where
\begin{align}
    c_{\bm{k}I} = \begin{pmatrix} c_{-\bm{k}I} \\ c_{+\bm{k}I} \end{pmatrix}, && c^\dagger_{\bm{k}I} = \begin{pmatrix} c^\dagger_{-\bm{k}I} & c^\dagger_{+\bm{k}I} \end{pmatrix}.
\end{align}
Here, we introduced the index $I \in \{K_1, K_1', K_2, K_2'\}$. It denotes at which Dirac cone each state is "based." Note that $\eta$ can be extracted directly from $I$.
The effective electron Hamiltonian for each of the cones is
\begin{align}
    \tilde{H}_n^{I, s}(\bm{k}) =  \varepsilon_{n\bm{k}} c^{\dagger}_{n\bm{k}I s}c_{n\bm{k}I s}.
\end{align}
where $n$ denotes the energy levels $n=\pm$ and we introduced the electron spin $s$.

\section{Interfacial electron-magnon coupling\label{S-d}}
In this section, we consider the s-d coupling between the ferromagnetic layers and the electrons in the graphene layers. We consider a local coupling
\begin{align}
    H_{\mathrm{int}} = J_{\mathrm{s-d}} \int \mathrm{d}^2\bm{r} \bm{m}(\bm{r})\cdot \bm{s}(\bm{r}),
    \label{startSD}
\end{align}
where $\bm{m}$ is the magnetization in the ferromagnet and $\bm{s}$ is the spin of the itinerant electrons. We rewrite the spin operators in terms of raising and lowering operators
\begin{align}
    s_x=\frac{1}{2}(s^+ + s^-) && s_x=\frac{1}{2i}(s^+ - s^-).
\end{align}
As a result, we find
\begin{align}
    H_{\mathrm{int}} = J_{\mathrm{s-d}} \int \mathrm{d}^2\bm{r} \left[ \frac{1}{2}\left(m^+s^- + m^-s^+\right) + m_z s_z \right].
\end{align}
We use the same Holstein-Primakoff transformation as in Eqs. \eqref{HPtop} and \eqref{HPbottom} with a subsequent Fourier transform to find
\begin{subequations}
\begin{align}
    H^T_{\mathrm{int}} =  \sum_{\bm{q},\bm{k}}J_{\mathrm{s-d}} \frac{\sqrt{M}}{\sqrt{2\mathcal{V}}}\left(a_{\bm{q}}\psi^{\dagger}_{a,\bm{k}+\bm{q},\uparrow}\psi_{a,\bm{k},\downarrow} + a^\dagger_{\bm{q}}\psi^{\dagger}_{a,\bm{k}-\bm{q},\downarrow}\psi_{a,\bm{k},\uparrow}\right) + \sum_{\bm{k}}J_{\mathrm{s-d}}M\left(\psi^{\dagger}_{a,\bm{k},\uparrow}\psi_{a,\bm{k},\uparrow}-\psi^{\dagger}_{a,\bm{k},\downarrow}\psi_{a,\bm{k},\downarrow}\right),
    \label{hamA}
\end{align}
and
\begin{align}
    H^B_{\mathrm{int}} =  \sum_{\bm{q},\bm{k}}J_{\mathrm{s-d}} \frac{\sqrt{M}}{\sqrt{2\mathcal{V}}}\left(b^{\dagger}_{\bm{q}}\psi^{\dagger}_{b,\bm{k}-\bm{q},\uparrow}\psi_{b,\bm{k},\downarrow} + b_{\bm{q}}\psi^{\dagger}_{b,\bm{k}+\bm{q},\downarrow}\psi_{b,\bm{k},\uparrow}\right) - \sum_{\bm{k}}J_{\mathrm{s-d}}M\left(\psi^{\dagger}_{b,\bm{k},\uparrow}\psi_{b,\bm{k},\uparrow}-\psi^{\dagger}_{b,\bm{k},\downarrow}\psi_{b,\bm{k},\downarrow}\right),
    \label{hamB}
\end{align}
\end{subequations}
for the top (T) and bottom layer (B). Here, $\psi^{\dagger}_{a,\bm{k},s}$ and $\psi^{\dagger}_{b,\bm{k},s}$ are operators of electron states located in the top and bottom layer, respectively, with momentum $\bm{k}$ and spin $s$. We use the layer indices $a$ and $b$ to emphasize that the states are not associated with particular points in the Brillouin zone. The low-energy electron states, however, are located close to the Dirac cones. We now project the Hamiltonians in Eq. \eqref{hamA} and \eqref{hamB} on the electron states given in Eq. \eqref{bases}.
Furthermore, we restrict our discussion to intravalley scattering. Thus, the top layer Hamiltonian can be written as
\begin{align}
\begin{split}
    H^T_{\mathrm{int}} =  J_{\mathrm{s-d}} \frac{\sqrt{M}}{\sqrt{2\mathcal{V}}}\sum_{\bm{q},\bm{k}}a_{\bm{q}}\left(\left(\psi^{{K}_{1}\dagger}_{\uparrow,\bm{k}+\bm{q}}\psi^{{K}_{1}}_{\downarrow,\bm{k}} + \psi^{{K'}_{1}\dagger}_{\uparrow,\bm{k}+\bm{q}}\psi^{{K'}_{1}}_{\downarrow,\bm{k}}\right)
    + \sum^3_{j=1} \left(\psi^{K_{2}\dagger}_{\uparrow,\bm{k}+\bm{q}_j+\bm{q}}\psi^{K_{2}}_{\downarrow,\bm{k}+\bm{q}_j} + \psi^{{K'}_{2}\dagger}_{\uparrow,\bm{k}-\bm{q}_j+\bm{q}}\psi^{{K'}_{2}}_{\downarrow,\bm{k}-\bm{q}_j}\right)\right) +h.c. \\+ J_{\mathrm{s-d}}M\sum_{s,\bm{k}}s\biggl(\left(\psi^{{K}_{1}\dagger}_{s,\bm{k}}\psi^{{K}_{1}}_{s,\bm{k}}+\psi^{{K'}_{1}\dagger}_{s,\bm{k}}\psi^{{K'}_{1}}_{s,\bm{k}}\right)
    +\sum^3_{j=1} \left(\psi^{K_{2}\dagger}_{s,\bm{k}+\bm{q}_j}\psi^{K_{2}}_{s,\bm{k}+\bm{q}_j}+\psi^{{K'}_{2}\dagger}_{s,\bm{k}-\bm{q}_j}\psi^{{K'}_{2}}_{s,\bm{k}-\bm{q}_j}\right)\biggr).
    \label{sd80Explicit}
\end{split}
\end{align}
Here, the electron spin index $s=\pm 1$ is positive for spin up $\uparrow$ and negative for spin down $\downarrow$.
We now transform the coupling and project the Hamiltonian onto the low-energy states given in Eq. \eqref{eigenstatesSummary}. 
In this case, the coupling simplifies to
\begin{align}
\begin{split}
    H^T_{\mathrm{int}} =  \frac{J_{\mathrm{s-d}}}{1+6\alpha^2} \frac{\sqrt{M}}{\sqrt{2\mathcal{V}}}\sum_{\bm{q},\bm{k}}a_{\bm{q}}\biggl(\Psi^{K_1\dagger}_{\uparrow,\bm{k}+\bm{q}}\Psi^{K_1}_{\downarrow,\bm{k}} + \Psi^{K'_1\dagger}_{\uparrow,\bm{k}+\bm{q}}\Psi^{K'_1}_{\downarrow,\bm{k}} 
    + 6\alpha^2\left( \Psi^{K_2\dagger}_{\uparrow,\bm{k}+\bm{q}}\Psi^{K_2}_{\downarrow,\bm{k}}
   +\Psi^{K'_2\dagger}_{\uparrow,\bm{k}+\bm{q}}\Psi^{K'_2}_{\downarrow,\bm{k}}\biggr)\right)+h.c
   \\ 
   +\frac{J_{\mathrm{s-d}}M}{1+6\alpha^2} \sum_{s,\bm{k}}s\biggl(\Psi^{K_1 \dagger}_{s,\bm{k}}\Psi^{K_1}_{s,\bm{k}}+\Psi^{K_1'\dagger}_{s,\bm{k}}\Psi^{K_1'}_{s,\bm{k}}
    +6\alpha^2\left(\Psi^{K_2\dagger}_{s,\bm{k}}\Psi^{K_2}_{s,\bm{k}} +\Psi^{K'_2\dagger}_{s,\bm{k}}\Psi^{K'_2}_{s,\bm{k}} \right)\biggr).
    \label{sdTop}
\end{split}
\end{align}
The bottom ferromagnet couples to the bottom graphene layer in a similar way. The explicit form is
\begin{align}
\begin{split}
    H^B_{\mathrm{int}} =  \frac{J_{\mathrm{s-d}}}{1+6\alpha^2} \frac{\sqrt{M}}{\sqrt{2\mathcal{V}}}\sum_{\bm{q}\bm{k}}b^{\dagger}_{\bm{q}}\biggl(\Psi^{K_2\dagger}_{\uparrow\bm{k}-\bm{q}}\Psi^{K_2}_{\downarrow\bm{k}} + \Psi^{K'_2\dagger}_{\uparrow\bm{k}-\bm{q}}\Psi^{K'_2}_{\downarrow\bm{k}} 
    + 6\alpha^2\left( \Psi^{K_1'\dagger}_{\uparrow\bm{k}-\bm{q}}\Psi^{K'_1}_{\downarrow\bm{k}}
   + \Psi^{K_1\dagger}_{\uparrow\bm{k}-\bm{q}}\Psi^{K_1}_{\downarrow\bm{k}}\right)\biggr)+h.c \\
   +\frac{J_{\mathrm{s-d}}M}{1+6\alpha^2} \sum_{s,\bm{k}}-s\biggl(\left(\Psi^{K_2\dagger}_{s,\bm{k}}\Psi^{K_2}_{s,\bm{k}}+\Psi^{K'_2\dagger}_{s,\bm{k}}\Psi^{K'_2}_{s,\bm{k}}\right)
    +6\alpha^2\left(\Psi^{K'_1\dagger}_{s,\bm{k}}\Psi^{K'_1}_{s,\bm{k}\uparrow} +\Psi^{K_1\dagger}_{s,\bm{k}}\Psi^{K_1}_{s,\bm{k}} \right)\biggr)
    \label{sdBottom}
\end{split}
\end{align}
The total s-d coupling is the sum of Eqs. \eqref{sdTop} and \eqref{sdBottom}. Note how the exchange splitting does not cancel out except at the special angle $\alpha^2=1/6$. However, this is not the magic angle $\alpha^2=1/3$. The coupling is expressed in terms of the low-energy eigenstates of the BM model. We transform the coupling using Eq. \eqref{UTransform}, to express the coupling in terms of band electrons.
To simplify the model, we consider coupling to the lower band only. In terms of lower band electron operators, the interfacial coupling is
\begin{align}
\begin{split}
    H^T_{\mathrm{s-d}} =  \frac{J_{\mathrm{s-d}}}{1+6\alpha^2} \frac{\sqrt{M}}{\sqrt{2\mathcal{V}}}\sum_{\bm{q},\bm{k}}a_{\bm{q}}\biggl(c_{\bm{k}+\bm{q}K_2' \uparrow}^\dagger \phi^{K\dagger}_{\bm{k}+\bm{q}} \phi^{K}_{\bm{k}} c_{\bm{k}K_2' \downarrow} + c_{\bm{k}+\bm{q}K_1' \uparrow}^\dagger \phi^{K'\dagger}_{\bm{k}+\bm{q}} \phi^{K'}_{\bm{k}} c_{\bm{k}K_1' \downarrow} 
    \\ + 6\alpha^2 \left(c_{\bm{k}+\bm{q}K_2 \uparrow}^\dagger \phi^{K\dagger}_{\bm{k}+\bm{q}} \phi^{K}_{\bm{k}} c_{\bm{k}K_2 \downarrow}
    c_{\bm{k}+\bm{q}K_1' \uparrow}^\dagger \phi^{K'\dagger}_{\bm{k}+\bm{q}} \phi^{K'}_{\bm{k}} c_{\bm{k}K_1' \downarrow}\right)\biggr)+h.c.
    \\ +\frac{J_{\mathrm{s-d}}M}{1+6\alpha^2} \sum_{s,\bm{k}}s\biggl(\left(c_{\bm{k}K_1 s}^\dagger c_{\bm{k}K_1 \uparrow}+c_{\bm{k}K_1' s}^\dagger c_{\bm{k}K_1' s}\right) 
    +6\alpha^2\left(c_{\bm{k}K_2 s}^\dagger c_{\bm{k}K_2 s} +c_{\bm{k}K_2' s}^\dagger c_{\bm{k}K'_2 s}\right)\biggr).
    \label{sdTopBand}
\end{split}
\end{align}
The $\phi$-factors can be written as
\begin{align}
    \phi_{\bm{k}+\bm{q}}^{\eta\dagger}\phi^{\eta}_{\bm{k}} = \frac{1}{2}\left(e^{i\eta(\varphi_{\bm{k}}-\varphi_{\bm{k}+\bm{q}})}+1\right), && \phi_{\bm{k}-\bm{q}}^{\eta\dagger}\phi^{\eta}_{\bm{k}} = \frac{1}{2}\left(e^{i\eta(\varphi_{\bm{k}}-\varphi_{\bm{k}-\bm{q}})}+1\right).
\end{align}
The analogous coupling for the bottom layer coupling is
\begin{align}
\begin{split}
    H^B_{\mathrm{s-d}} =  \frac{J_{\mathrm{s-d}}}{1+6\alpha^2} \frac{\sqrt{M}}{\sqrt{2\mathcal{V}}}\sum_{\bm{q}\bm{k}}b^{\dagger}_{\bm{q}}\biggl(c_{\bm{k}-\bm{q}K_2 \uparrow}^\dagger \phi^{K\dagger}_{\bm{k}-\bm{q}} \phi^{K}_{\bm{k}} c_{\bm{k}K_2 \downarrow} + c_{\bm{k}-\bm{q}K'_2 \uparrow}^\dagger \phi^{K'\dagger}_{\bm{k}-\bm{q}} \phi^{K'}_{\bm{k}} c_{\bm{k}K'_2 \downarrow} 
    \\ + 6\alpha^2 \left(c_{\bm{k}-\bm{q}K_1 \uparrow}^\dagger \phi^{K\dagger}_{\bm{k}-\bm{q}} \phi^{K}_{\bm{k}} c_{\bm{k}K_1 \downarrow}
   +c_{\bm{k}-\bm{q}K_1' \uparrow}^\dagger \phi^{K'\dagger}_{\bm{k}-\bm{q}} \phi^{K'}_{\bm{k}} c_{\bm{k}K_1' \downarrow}\right)\biggr)+h.c.
   \\
   +\frac{J_{\mathrm{s-d}}M}{1+6\alpha^2} \sum_{s,\bm{k}}(-s)\biggl(\left(c_{\bm{k}K_2 s}^\dagger c_{\bm{k}K_2 s}+c_{\bm{k}K_2' s}^\dagger c_{\bm{k}K_2' s}\right)
    +6\alpha^2\left(c_{\bm{k}K'_1 s}^\dagger c_{\bm{k}K'_1 s}+ c_{\bm{k}K_1 s}^\dagger c_{\bm{k}K_1 s} \right)\biggr).
    \label{sdBottomBand}
\end{split}
\end{align}
From Eq. \eqref{sdTopBand} and \eqref{sdBottomBand}, we can extract the exchange splitting $\Delta_{ss}$ and electron-magnon coupling constant $g$ introduced in the main text.

\section{Effective interaction\label{Effective interaction}}
The electron-magnon coupling can give rise to an effective electron-electron interaction. We consider the system Hamiltonian
\begin{align}
    H=H_{\mathrm{e}} + H_{\mathrm{m}} + H_{\mathrm{em}}.
\end{align}
The electron Hamiltonian is
\begin{subequations}
\begin{align}
    H_{\mathrm{e}}=& \sum_{\bm{k},I, s}\varepsilon_{\bm{k}}c_{\bm{k}I s}^\dagger c_{\bm{k}I s} + \Delta_{ss}\sum_{\bm{k},I,s }ls\left(c_{\bm{k}I s}^{\dagger}c_{\bm{k}I s}\right),
    \label{electronHamiltonian}
\end{align}
where we introduced the layer index $l$. It can be extracted directly from the index $I$, and takes the value $l=1$ for $I \in \{K_1, K'_1\}$ and $l=-1$ for $I \in \{K_2, K'_2\}$. Hence, top-layer spin-up electrons experience the same exchange spin shift as a bottom-layer spin-down electron due to the identical but anti-parallel ferromagnets. The magnitude of the spin splitting is 
\begin{align}
    \Delta_{ss}=\frac{J_{\mathrm{s-d}}M(1-6\alpha^2)}{1+6\alpha^2}.
    \label{SpinSplitting}
\end{align}
The electron dispersion $\varepsilon_{\bm{k}}$ is that of the lower band.
The magnon Hamiltonian is
\begin{align}
    H_{\mathrm{m}} =& \sum_{\bm{q}}\omega_{\bm{q}} \left(a_{\bm{q}}^\dagger a_{\bm{q}} + b_{\bm{q}}^\dagger b_{\bm{q}}\right),
\end{align}
and the electron-magnon coupling is
\begin{align}
\begin{split}
    H_{\mathrm{em}} =& \sum_{\bm{q}\bm{k}I}g_{I,\bm{k},\bm{q}}^{a} \left(c_{\bm{k}+\bm{q}I\uparrow}^\dagger c_{\bm{k}I\downarrow} \right)a_{\bm{q}} + (g_{I,\bm{k},\bm{q}}^{a})^* \left(c_{\bm{k}I\downarrow}^\dagger c_{\bm{k}+\bm{q}I\uparrow} \right)a_{\bm{q}}^{\dagger} \\ &+ g_{I,\bm{k},-\bm{q}}^{b} \left(c_{\bm{k}-\bm{q}I\uparrow}^\dagger c_{\bm{k}I\downarrow} \right)b^{\dagger}_{\bm{q}} + (g_{I,\bm{k},-\bm{q}}^{b})^* \left(c_{\bm{k}I\downarrow}^\dagger c_{\bm{k}-\bm{q}I\uparrow} \right)b_{\bm{q}}.
\end{split}
\end{align}
\end{subequations}
The coupling coefficients depend on the layer at which the electron state is based and the chirality of the Dirac cone $I$. To account for this, we consider two disjoint subsets of $I$. That is, we denote $\{K_1, K'_1\}$ by $I_1$ and $\{K_2, K'_2\}$ by $I_2$. The coupling constants are then
\begin{subequations}
\begin{align}
    g_{I_1,\bm{k},\bm{q}}^{a} =& g_{I_2,\bm{k},\bm{q}}^{b} = g_0\phi^{\eta\dagger}_{\bm{k}+\bm{q}} \phi^{\eta}_{\bm{k}}, \\
    g_{I_2,\bm{k},\bm{q}}^{a} =& g_{I_1,\bm{k},\bm{q}}^{b} = 6\alpha^2 g_0\phi^{\eta\dagger}_{\bm{k}+\bm{q}} \phi^{\eta}_{\bm{k}},
\end{align}
\end{subequations}
where
\begin{align}
    g_0=\frac{J_{\mathrm{s-d}}}{1+6\alpha^2} \frac{\sqrt{M}}{\sqrt{2\mathcal{V}}}.
\end{align}
The $a$ and $b$ superscripts denote the top and bottom layers, respectively. The $\phi$-factors are defined in Eq. \eqref{phiFactors}.

We relabel the indices of $H_{\mathrm{em}}$ and use $(g_{\bm{k}+\bm{q},-\bm{q}}^{I})^*=g_{\bm{k}, \bm{q}}^{I}$. The relabeling simplifies the electron-magnon coupling to
\begin{align}
\begin{split}
    H_{\mathrm{em}} =& \sum_{\bm{q},\bm{k},I}g_{I,\bm{k},\bm{q}}^{a} \left(c_{\bm{k}+\bm{q}I\uparrow}^\dagger c_{\bm{k}I\downarrow} \right)a_{\bm{q}} + g_{I,\bm{k},\bm{q}}^{a} \left(c_{\bm{k}+\bm{q}I\downarrow}^\dagger c_{\bm{k}I\uparrow} \right)a_{-\bm{q}}^{\dagger} \\ &+ g_{I,\bm{k},\bm{q}}^{b} \left(c_{\bm{k}+\bm{q}I\uparrow}^\dagger c_{\bm{k}I\downarrow} \right)b^{\dagger}_{-\bm{q}} + g_{I,\bm{k},\bm{q}}^{b} \left(c_{\bm{k}+\bm{q}I\downarrow}^\dagger c_{\bm{k}I\uparrow} \right)b_{\bm{q}}.
\end{split}
\end{align}

We split the Hamiltonian into two parts to perform the Schrieffer-Wolff transformation \cite{schrieffer1966relation}.
Let $H_0 = H_{\mathrm{e}} + H_{\mathrm{m}}$ and $H_1 = H_{\mathrm{em}}$ such that we have $H = H_0 + \eta H_1$. The small expansion parameter $\eta$ should not be confused with the valley index. We consider the canonical transformation
\begin{align}
    H'=e^{-\eta S} H e^{\eta S}
\end{align}
and expand to find
\begin{align}
\begin{split}
    H' = H_0 + \eta(H_1 + \left[H_0, S\right]) + \eta^2\left[H_1, S\right] + \frac{\eta^2}{2}\left[\left[H_0, S\right], S\right] + \mathcal{O}(\eta^3).
\end{split}
\end{align}
To eliminate the linear term, we require
\begin{align}
    H_1 + \left[H_0, S\right] = 0.
    \label{condition}
\end{align}
To that end, we make the ansatz
\begin{align}
\begin{split}
    S =& \sum_{\bm{q},\bm{k},I}g_{I,\bm{k},\bm{q}}^{a} x^a_{\bm{k}, \bm{q},I}a_{\bm{q}}\left(c_{\bm{k}+\bm{q}I\uparrow}^\dagger c_{\bm{k}I\downarrow} \right) + g_{I,\bm{k},\bm{q}}^{a} y^a_{\bm{k}, \bm{q},I}a_{-\bm{q}}^{\dagger}\left(c_{\bm{k}+\bm{q}I\downarrow}^\dagger c_{\bm{k}I\uparrow} \right) \\&+ g_{I,\bm{k},\bm{q}}^{b}y^b_{\bm{k}, \bm{q},I}b^{\dagger}_{-\bm{q}} \left(c_{\bm{k}+\bm{q}I\uparrow}^\dagger c_{\bm{k}I\downarrow} \right) + g_{I,\bm{k},\bm{q}}^{b}x^b_{\bm{k}, \bm{q},I} b_{\bm{q}}\left(c_{\bm{k}+\bm{q}I\downarrow}^\dagger c_{\bm{k}I\uparrow} \right).
\end{split}
\end{align}
We evaluate Eq. \eqref{condition} and find
\begin{align}
\begin{split}
     x^a_{\bm{k}, \bm{q},I} = x^b_{\bm{k}, \bm{q},I} =   \frac{1}{\varepsilon_{\bm{k}} -\varepsilon_{\bm{k}+\bm{q}} + \omega_{\bm{q}}} \\
    y^a_{\bm{k}, \bm{q},I} = y^b_{\bm{k}, \bm{q},I} =  \frac{1}{\varepsilon_{\bm{k}} -\varepsilon_{\bm{k}+\bm{q}} - \omega_{\bm{q}}},
\end{split}
\end{align}
where we used that $\omega_{-\bm{q}}=\omega_{\bm{q}}$. The parameters $x$ and $y$ are independent of the index $I$ to second order in the small coupling parameter $J_{\mathrm{s-d}}$.
The effective interaction is
\begin{align}
    H_{\mathrm{eff}} = \frac{\eta^2}{2}\left[H_1, S\right].
\end{align}
The commutator $\left[H_1, S\right]$ has the form $\left[Aa, Bb\right]$ with $A,B\propto c^\dagger c$ containing products of fermion operators and $a,b \propto xa + ya^\dagger$ containing sums of boson operators. The general relationship
\begin{align}
    \left[Aa, Bb\right]=AB[a,b] + \left[A, B \right]ab - \left[A,B\right][a, b]
    \label{GenRel}
\end{align}
shows that there are three types of contributions. The last term represents a one-body electron operator, which can be shown to vanish \cite{schrieffer1966relation}. The second term describes an effective coupling of an electron to two magnons. We are interested in the effective electron-electron interaction described by the first term. The magnon operators of distinct layers commute
\begin{align}
    \left[a_{\bm{q}} + b^{\dagger}_{\bm{q}}, a^{\dagger}_{\bm{q}} + b_{\bm{q}}\right] = \left[a_{\bm{q}}, a^{\dagger}_{\bm{q}}\right] +\left[b^{\dagger}_{\bm{q}}, b_{\bm{q}}\right],
\end{align}
such that we can treat each layer independently to get
\begin{align}
\begin{split}
    H_{\mathrm{eff}}=\sum_{\bm{q},\bm{k},\bm{k}',I,I'} \left(g_{I,\bm{k},\bm{q}}^{a}g_{I',\bm{k}',-\bm{q}}^{a} + g_{I,\bm{k},\bm{q}}^{b}g_{I',\bm{k}',-\bm{q}}^{b}\right) (y_{\bm{k}', -\bm{q}}-x_{\bm{k}, \bm{q}})c_{\bm{k}+\bm{q}I\uparrow}^\dagger c_{\bm{k}I\downarrow} c_{\bm{k}'-\bm{q}I'\downarrow}^\dagger c_{\bm{k}'I'\uparrow}.
\end{split}
\end{align}

\subsection{Effective BCS type pairing}

We consider a BCS-type pairing. Hence, we let $\bm{k}'=-\bm{k}$. In this case,
\begin{align}
    y_{-\bm{k}, -\bm{q}} - x_{\bm{k}, \bm{q}} = \frac{2\omega_{\bm{q}}}{(\varepsilon_{\bm{k}}-\varepsilon_{\bm{k}+\bm{q}})^2 - \omega_{\bm{q}}^2},
\end{align}
where we used that $\varepsilon_{-\bm{k}-\bm{q}} = \varepsilon_{\bm{k}+\bm{q}}$.
Now, we relabel $\bm{k}+\bm{q}\rightarrow \bm{k}'$ and subsequently $\bm{k}\rightarrow -\bm{k}$ to get
\begin{align}
    H_{\mathrm{eff}}=\sum_{\bm{k}\bm{k}'II'} V_{\mathrm{eff}}({\bm{k},\bm{k}'}, I,I')\left(c_{\bm{k}'I\uparrow}^\dagger c_{-\bm{k}'I'\downarrow}^\dagger c_{-\bm{k}I\downarrow}  c_{\bm{k}I'\uparrow}\right).
    \label{effectiveInteraction1}
\end{align}
The effective interaction is
\begin{align}
\begin{split}
    V_{\mathrm{eff}}({\bm{k},\bm{k}'}, I,I') =  \frac{2\omega_{\bm{k'+k}}\left(\sum_{j\in \{a,b\}}g_{I,-\bm{k},\bm{k}'+\bm{k},}^{j} g_{I',-\bm{k},\bm{k}'+\bm{k}}^{j}\right)}{\omega^2_{\bm{k}'+\bm{k}}-(\varepsilon_{\bm{k}'}-\varepsilon_{\bm{k}})^2}.
    \label{effectiveInteractionExplicit}
\end{split}
\end{align}
Eq. \eqref{effectiveInteraction1} governs the effective interaction between electrons based at Dirac cones $I$ and $I'$. Of the multiple possible channels, only a few are suitable for Cooper pair formation. We denote the partner of $I$ by $\bar{I}$ and consider the pairs
\begin{align}
    \left\{I = K_1,\bar{I} = K_2\right\}, &&
    \left\{I = K'_1, \bar{I} = K'_2\right\}.
    \label{PairingChannel}
\end{align}
Note how the electron states are based in distinct layers. For this reason, they are not spin split.

For this particular choice of electron pairs, the effective interaction simplifies to
\begin{align}
\begin{split}
    V_{\mathrm{eff}}({\bm{k},\bm{k}'}, I,\bar{I}) =  \frac{\omega_{\bm{k'+k}}}{\omega^2_{\bm{k}'+\bm{k}}-(\varepsilon_{\bm{k}'}-\varepsilon_{\bm{k}})^2}\frac{12\alpha^2J^2_{\mathrm{s-d}}}{(1+6\alpha^2)^2} \frac{M}{\mathcal{V}}w^{\eta\eta'}_{\bm{k}\bm{k}'},
    \label{effectiveInteraction}
\end{split}
\end{align}
where
\begin{align}
    w^{\eta\eta'}_{\bm{k}\bm{k}'}=\phi_{\bm{k}'}^{\eta \dagger}\phi_{-\bm{k}}^{\eta}\phi_{\bm{k}'}^{\eta' \dagger}\phi_{-\bm{k}}^{\eta'}.
\end{align}
The angular dependence of the effective interaction is determined by the magnon-dispersion relation $\omega_{\bm{k}+\bm{k}'}$ and $w^{\eta\eta'}_{\bm{k}\bm{k}'}$. To simplify the angular dependence, we approximate the magnon dispersion to be constant $\omega_{\bm{k}+\bm{k}'}=\omega_M$ and expand $w^{\eta\eta'}_{\bm{k}\bm{k}'}$
\begin{align}
\begin{split}
     w^{\eta\eta'}_{\bm{k}\bm{k}'}=&\phi_{\bm{k}'}^{\eta \dagger}\phi_{-\bm{k}}^{\eta}\phi_{\bm{k}'}^{\eta' \dagger}\phi_{-\bm{k}}^{\eta'} \\=&
     e^{i\frac{\eta+\eta'}{2}(\phi_{\bm{k}}-\phi_{\bm{k}'})}\left(\frac{1}{4}\left( e^{i(\phi_{\bm{k}}-\phi_{\bm{k}'})} + e^{-i(\phi_{\bm{k}}-\phi_{\bm{k}'})}\right) - \frac{1}{2}\right)
     \\w^{\eta\eta}_{\bm{k}\bm{k}'}=&e^{i\eta(\phi_{\bm{k}}-\phi_{\bm{k}'})}\left(\frac{1}{4}\left( e^{i(\phi_{\bm{k}}-\phi_{\bm{k}'})} + e^{-i(\phi_{\bm{k}}-\phi_{\bm{k}'})}\right) - \frac{1}{2}\right)
     \label{angularDep}
\end{split}
\end{align}
In the last line, we used that $I$ and $\bar{I}$ have the same chirality.
Eq. \eqref{angularDep} suggests that the effective interaction is separable with respect to angular character.

First, let $(I, \bar{I}) = (K_1, K_2)$ such that $\eta=1$. For this pair,
\begin{subequations}
\begin{align}
    w^{\eta\eta}_{\bm{k}\bm{k}'}=&\left(\frac{1}{4}\left(e^{2i(\phi_{\bm{k}}-\phi_{\bm{k}'})} + 1\right) - \frac{1}{2}e^{i(\phi_{\bm{k}}-\phi_{\bm{k}'})}\right).
\end{align}
The two first terms correspond to $d$- and $s$-wave characters with respect to rotations in the plane. In the low-frequency limit $(\varepsilon_{\bm{k}'}-\varepsilon_{\bm{k}})^2\rightarrow 0$, these channels render a repulsive interaction. Conversely, the last term yields an attractive interaction with a $p$-wave character.
Second, we consider $(I, \bar{I}) = (K'_1, K'_2)$ such that $\eta=-1$. This pair yields the coupling
\begin{align}
    w^{\eta\eta}_{\bm{k}\bm{k}'}=&\left(\frac{1}{4}\left(e^{-2i(\phi_{\bm{k}}-\phi_{\bm{k}'})} + 1\right) - \frac{1}{2}e^{-i(\phi_{\bm{k}}-\phi_{\bm{k}'})}\right).
\end{align}
\end{subequations}
with an attractive $p$-wave interaction.
The pairing channels and effective interactions are related through time-reversal symmetry.

\section{BCS gap equation\label{BCS gap equation}}
In this section, we consider the effective electron Hamiltonian. 
\begin{align}
    H =\sum_{\bm{k}Is} \varepsilon_{\bm{k}Is} c_{\bm{k}I s}^\dagger c_{\bm{k}I s} + \sum_{\bm{k}\bm{k}'I} V_{\mathrm{eff}}({\bm{k},\bm{k}'}, I,\bar{I})c_{\bm{k}I\uparrow}^\dagger c_{-\bm{k}\bar{I}\downarrow}^\dagger c_{-\bm{k}'I\downarrow}  c_{\bm{k}'\bar{I}\uparrow}.
\end{align}
Here, the first term corresponds to the electron dispersion Eq. \eqref{electronHamiltonian}, including the exchange splitting effect. The second term corresponds to the effective interaction in Eq. \eqref{effectiveInteraction1} for the pairs defined in Eq. \eqref{PairingChannel}.

Following the conventional BCS approach, we introduce two hermitian conjugate gap parameters. We define
\begin{subequations}
\begin{align}
    \Delta_{I\bm{k}} = -\sum_{\bm{k'}}V_{\bm{k}\bm{k}'I\bar{I}}\left< c_{-\bm{k}'I\downarrow}  c_{\bm{k}'\bar{I}\uparrow}\right>, \\
    \Delta_{I\bm{k}}^\dagger = -\sum_{\bm{k}'}V_{\bm{k}'\bm{k}\bar{I}I}\left<c_{\bm{k}'\bar{I}\uparrow}^\dagger c_{-\bm{k}'I\downarrow}^\dagger\right>.
    \label{DefinitionGap}
\end{align}
\end{subequations}
Here, $V_{\bm{k}\bm{k}'I\bar{I}}$ is shorthand notation for $V_{\mathrm{eff}}({\bm{k},\bm{k}'}, I,\bar{I})$.
The resulting mean-field Hamiltonian is
\begin{align}
\begin{split}
    H_{\mathrm{BCS}} = \sum_{\bm{k}Is} \varepsilon_{\bm{k}Is} c_{\bm{k}I s}^\dagger c_{\bm{k}I s}  - \sum_{\bm{k}I} \Delta_{I\bm{k}}^\dagger c_{-\bm{k}\bar{I}\downarrow}  c_{\bm{k}I\uparrow} - \sum_{\bm{k}I}\Delta_{I\bm{k}} c_{\bm{k}I\uparrow}^\dagger c_{-\bm{k}\bar{I}\downarrow}^\dagger - \mathrm{const}.
    \label{BCSHam}
\end{split}
\end{align}
To diagonalize this Hamiltonian, we introduce new fermionic operators through the Bogoliubov transformation
\begin{align}
    \begin{pmatrix} \gamma_{\bm{k}I\uparrow} \\ \gamma_{-\bm{k}\bar{I}\downarrow}^{\dagger}\end{pmatrix} = \begin{pmatrix} u_{I\bm{k}}^* & - v_{I\bm{k}} \\ v_{I\bm{k}}^* & u_{I\bm{k}}\end{pmatrix}\begin{pmatrix} c_{\bm{k}I\uparrow} \\ c_{-\bm{k}\bar{I}\downarrow}^\dagger\end{pmatrix}.
\end{align}
We require the $\gamma$-operators to satisfy fermionic anticommutation relations
\begin{align}
    \left\{\gamma_{\bm{k}I\uparrow}, \gamma_{\bm{k}I\uparrow}^{\dagger}\right\} = \gamma_{\bm{k}I\uparrow}\gamma_{\bm{k}I\uparrow}^{\dagger} + \gamma_{\bm{k}I\uparrow}^{\dagger}\gamma_{\bm{k}I\uparrow}
\end{align}
which is equivalent to
\begin{align}
    \abs{u_{I\bm{k}}}^2 + \abs{v_{I\bm{k}}}^2 = 1.
    \label{determinant}
\end{align}
Furthermore, we choose $u_{I\bm{k}}$ and $v_{I\bm{k}}$ to diagonalize the Hamiltonian with respect to the quasiparticle $\gamma$ operators.
We find a special solution by choosing
\begin{align}
    \Delta_{I\bm{k}} = \abs{\Delta_{I\bm{k}}}e^{i\phi_{\bm{k}}}, && u_{I\bm{k}}=\abs{u_{I\bm{k}}}, && v_{I\bm{k}} = \abs{v_{I\bm{k}}}e^{i\phi_{\bm{k}}}.
\end{align}
For this choice, the equation simplifies to
\begin{align}
    2\varepsilon_{\bm{k}I\uparrow}\abs{u_{I\bm{k}}}\abs{v_{I\bm{k}}} + \abs{\Delta_{I\bm{k}}}\left(\abs{v_{I\bm{k}}}^2 - \abs{u_{I\bm{k}}}^2\right)=0.
\end{align}
Here, we used the identity $\varepsilon_{-\bm{k}Is} = \varepsilon_{-\bm{k}\bar{I}\bar{s}}$, with $\bar{s}$ being the opposite spin of $s$.
By squaring the equation and employing Eq.  \eqref{determinant}, we find the relations
\begin{subequations}
\begin{align}
    \abs{u_{I\bm{k}}}^2 = \frac{1}{2}\left(1+\frac{\varepsilon_{\bm{k}I\uparrow}}{\sqrt{\varepsilon_{\bm{k}I\uparrow}^2 + \abs{\Delta_{I\bm{k}}}^2}}\right), &&
    \abs{v_{I\bm{k}}}^2 = \frac{1}{2}\left(1-\frac{\varepsilon_{\bm{k}I\uparrow}}{\sqrt{\varepsilon_{\bm{k}I\uparrow}^2 + \abs{\Delta_{I\bm{k}}}^2}}\right).
\end{align}
and
\begin{align}
    u_{I\bm{k}}v_{I\bm{k}} = \frac{\Delta_{I\bm{k}}}{2\sqrt{\varepsilon_{\bm{k}I\uparrow}^2 + \abs{\Delta_{I\bm{k}}}^2}}.
\end{align}
\end{subequations}
The diagonal form of the Hamiltonian in Eq. \eqref{BCSHam} is
\begin{align}
\begin{split}
    H_{\mathrm{BCS}} = \sum_{\bm{k}I}  \left(\sqrt{\varepsilon_{\bm{k}I\uparrow}^2 + \abs{\Delta_{I\bm{k}}}^2}\right)\gamma_{\bm{k}I\uparrow}^\dagger \gamma_{\bm{k}I\uparrow} +  \left(\sqrt{\varepsilon_{\bm{k}I\uparrow}^2 + \abs{\Delta_{I\bm{k}}}^2}\right)\gamma_{-\bm{k}\bar{I}\downarrow}^\dagger \gamma_{-\bm{k}\bar{I}\downarrow}.
\end{split}
\end{align}
The quasiparticle dispersion is spin-degenerate for this specific choice of electron pairs.

To establish the superconducting gap equation, we evaluate the expectation values in Eq. \eqref{DefinitionGap} in terms of the quasiparticle operators. We find the gap equation to be
\begin{align}
\begin{split}
    \Delta_{I\bm{k}}
    = -\sum_{\bm{k}'}V_{\bm{k}\bm{k}'I\bar{I}}\frac{\Delta_{I\bm{k}'}}{2E_{\bm{k}'I}}\tanh{\left(\frac{\beta E_{\bm{k}'I}}{2}\right)}.
\end{split}
\end{align}
The quasiparticle dispersion is given by
\begin{align}
    E_{\bm{k}I} = \sqrt{\left(\varepsilon_{\bm{k}} + \Delta^I_{ss}\right)^2 + \abs{\Delta_{I\bm{k}}}^2}
\end{align}
where the layer dependent exchange splitting is $\Delta^I_{ss} = \Delta_{ss} $ for $I\in \{K_1, K'_1\}$ and $\Delta^I_{ss} = -\Delta_{ss}$ for $I\in \{K_2, K'_2\}$.
We consider the gap equation for the attractive $p$-wave channel specifically. To that end, we let $V_{\mathrm{eff}}({\bm{k},\bm{k}'}, I,\bar{I}) = V e^{i\eta(\phi_{\bm{k}}-\phi_{\bm{k}'})}$ with
\begin{align}
    V = \frac{-6\alpha^2J^2_{\mathrm{s-d}}}{\omega_{M}(1+6\alpha^2)^2} \frac{M}{\mathcal{V}}
\end{align}
and $\Delta_{I\bm{k}} = \Delta_{I}e^{i\eta \phi_{\bm{k}}}$. To estimate the order of magnitude of the critical temperature, we consider the gap equation to the second order in the small coupling parameter $J_{\mathrm{s-d}}$. In this case, the critical temperature $T_c$ is given by the standard expression
\begin{align}
    T_c \approx \frac{1.13\omega_M}{k_B} e^{-\frac{1}{\lambda}}.
    \label{CriticalTemperature}
\end{align}
Here, $k_B$ is the Boltzmann constant and $\lambda = V N_D'$ where $N_D'$ is the electron density of states.

\section{Coulomb screening and candidate materials\label{Candidate materials}}
Twisted bilayer graphene serves as an excellent candidate for exhibiting magnon-mediated superconductivity. One of the reasons is the greatly enhanced electron density of states (DOS) close to the magic angles. Ref. \cite{carr2019exact} reports a DOS over $N_D = 100 \; \mathrm{eV}^{-1} \mathrm{nm}^{-2}$ close to the magic angle. This suggests a DOS per valley per spin and per layer of $N_D' = 12.5\; \mathrm{eV}^{-1} \mathrm{nm}^{-2}$.

The Coulomb interaction in momentum space is given by
\begin{align}
    V_C(q) = \frac{2\pi}{\epsilon}\frac{e^2}{q},
\end{align}
where $e$ is the electron charge, $\epsilon$ is the dielectric constant and $q$ is the wave vector. The repulsive interaction can be detrimental to the superconducting state. However, it is largely screened by the high DOS. Ref. \cite{Goodwin2019PRB} reports a twist-angle-dependent dielectric constant $\epsilon$. At the magic angle, they find $\epsilon > 250$ for long wavelengths. This gives rise to a significant screening effect. The Coulomb coupling strength for the relevant wavelengths is then
\begin{align}
    \mu = V_C(k_{\theta}) N_D' \approx 1. 
\end{align}
The Coulomb coupling strength exceeds the attractive magnon-mediated coupling strength. However, it is effectively frequency independent at the magnon frequency cut-off. To account for this, we employ the Morel-Anderson model \cite{MorelAnderson1962}. It effectively renormalizes the Coulomb coupling strength as
\begin{align}
    \mu^* = \frac{\mu}{1+\mu\ln{\frac{\omega_p}{\omega_M}}} \approx 0.13.
\end{align}
Here, we used the observed interband plasmon frequency $\omega_p \approx 200 \; \mathrm{meV}$ as the Coulomb interaction cut off \cite{hesp2021observation}. We used $\omega_M=0.3$ meV for the magnon frequency, justified by the ferromagnets used for critical temperature estimates. The effective interaction strength takes the value $\lambda^* = \lambda - \mu^*$. This shows that the Coulomb repulsion only weakly affects the superconducting state due to the TBG's enhanced screening properties.

The on-site interaction is a substantial part of the repulsive Coulomb interaction. This part has an $s$-wave symmetry. The $p$-wave symmetry superconducting state circumvents the on-site repulsion completely, such that the Coulomb repulsion has an even weaker effect. We will not consider the decomposition of the Coulomb effect in further detail because the treatment is already at a crude level.

Interfacial coupling between graphene and ferromagnets has been studied both theoretically and experimentally for numerous materials \cite{swartz2012integration,averyanov2018high,wei2016strong,wang2015proximity,farooq2019switchable,zhang2018strong,holmes2020exchange}. Here, we consider two specific materials to give an order of magnitude estimate for the critical temperature.

EuO is a ferromagnetic semiconductor with Curie temperature $T_c = 69$ K. It has an fcc unit cell with lattice constant $5.1$ Å. Hence, two magnetic Eu$^{2+}$ ions per unit cell, each with spin $S=7\mu_B$, are located at the interface and thus accessible for interfacial s-d coupling \cite{mauger1986magnetic}. EuO thin films can be deposited on graphene epitaxially \cite{swartz2012integration, averyanov2018high}. The induced exchange coupling is found to be $\Delta=36$ meV \cite{yang2013proximity}. Due to the long periodicity of the moirè structure, we consider coupling to long-wavelength magnons. Hence, we consider magnons with a wavenumber $q \lesssim k_\theta$. At this momentum, the magnon dispersion is cut off at $\omega_M \approx 0.15$ meV \cite{Dietrich1975PRB}. These parameters yield a coupling constant $\lambda^* \approx 0.8$ and a critical temperature of $T_c \approx 0.5$ K when taking Coulomb interaction into account.

CrI$_3$ is a van der Waals ferromagnet down to the monolayer limit \cite{huang2017layer}. The crystal has two magnetic ions per unit cell. Each magnetic ion carries a magnetic moment $S=3\mu_B$ \cite{mcguire2015coupling}. CrI$_3$ hosts two magnonic modes accessible for electron-magnon coupling. Their respective energies at momentum $\bm{q}=0$ are $0.3$ meV and $17$ meV \cite{cenker2021direct}. The magnon dispersion with respect to the momentum $k_{\theta}$ is negligible compared to the magnon gap. Hence, we use $\omega_M = 0.3$ meV. In a graphene-CrI$_3$ heterostructure, CrI$_3$ is theoretically found to induce an interfacial exchange splitting of $20$ meV \cite{holmes2020exchange}.
Considering CrI$_3$ as the ferromagnet, we find a coupling constant $\lambda^* \approx 0.4$ and critical temperature $T_c \approx 0.3$ K.

\end{document}